# Colossal intrinsic exchange bias in epitaxial $CoFe_2O_4/Al_2O_3$ thin films

*Detian Yang,[1]† Yu Yun,[1]†\* Arjun Subedi,[1] Nicholas E. Rogers,[2] David M. Cornelison,[2] Peter A. Dowben,[1] Xiaoshan Xu[1,3]\**

[1] Department of Physics and Astronomy, University of Nebraska, Lincoln, Nebraska 68588, USA
[2] Department of Physics, Astronomy and Materials Science, Missouri State University, Springfield, Missouri 65897, USA
[3] Nebraska Center for Materials and Nanoscience, University of Nebraska, Lincoln, Nebraska 68588, USA

†These authors contributed equally to this work.
\*Corresponding author. Email: yuyun@unl.edu and xiaoshan.xu@unl.edu

## Abstract

In this work, we demonstrate a massive intrinsic exchange bias (3 kOe) in epitaxial $CoFe_2O_4(111)$ thin films deposited on $Al_2O_3(0001)$ substrates. This exchange bias is indicative of intrinsic exchange or a ferromagnetic material combined with an antiferromagnet. The analysis of structure, magnetism and electronic states corroborate that there is an interfacial layer CoO between the $CoFe_2O_4(111)$ thin film and the $Al_2O_3(0001)$ substrate. The power-law thickness dependence of the intrinsic exchange bias verifies its interfacial origin. This work suggests interfacial engineering can be an effective route for achieving large exchange bias.

## Introduction

At the interface between a ferromagnetic or ferrimagnetic (FM) and an antiferromagnetic (AFM), the interfacial exchange interaction may favor the magnetization of the FM material in certain direction. This tendency to pin the magnetization in on direction results in a bias in the magnetic hysteresis, a phenomenon called exchange bias (EB) [1,2]. EB is fundamental to many magnetic storage and spintronic devices [2-4], and continues to be extensively studied both experimentally [2,5] and theoretically [5-7]. Although the consensus is that EB originates from the pinning of magnetic moment at the FM/AFM interface, the plethora interface parameters and measurement conditions complicates the construction of a general microscopic mechanism.

A type of EB, called "intrinsic" EB, has been especially intriguing since it occurs at the interface between an FM material and a non-magnetic material, without a nominal AFM layer. Intrinsic EB has been reported in a variety of heterostructures where FM thin films are epitaxially deposited on paramagnetic or diamagnetic substrates, such as $LaNiO_3/LaMnO_3$ superlattices [8], $La_{2/3}Sr_{1/3}MnO_3/LaSrAlO_4$ [9], $SrRuO_3/LaAlO_3$ [10], Fe/MgO [11]. All proposed mechanisms suggest formation of interfacial layers with distinctly different magnetic ordering that can pin the FM magnetization. For $LaNiO_3/LaMnO_3$ superlattices, EB comes from the induced magnetization associated with charge transfer at the interface [8]. In $La_{2/3}Sr_{1/3}MnO_3/LaSrAlO_4$, a strain-induced $LaSrMnO_4$-based spin glass layer forms [9]. While for $SrRuO_3/LaAlO_3$, an AFM $SrRuO_3$ interfacial layer is seen [10]. With Fe/MgO, FeO patches form at the interface due to oxygen diffusion from the substrate, is regarded as the AFM layer. These interfacial layers are believed to pin the FM magnetization.

On the other hand, since intrinsic EB relies on formation of interfacial layers, the heterostructures where intrinsic EB were initially discovered [8-11], are actually not expected to have large effects, because of the film/substrate structural similarity. In other words, intrinsic EB, like other emergent interfacial phenomena, is expected to be enhanced in epitaxial heterostructures of large mismatch of film/substrate structures; this is why a large intrinsic EB around 2 kOe was observed in hexagonal $Cr_2Te_3$ thin films deposited on a zinc-blende phase CdTe buffer layer [12]. In this regard, $CoFe_2O_4$ (CFO) thin films deposited on $Al_2O_3$ substrates, appear to be a promising heterostructure for achieving even larger intrinsic EB.

CFO is a ferrimagnetic insulator of large magnetocrystalline anisotropy, moderate magnetization, superior mechanic hardness and excellent physical and chemical stability, which has been widely studied [13,14] and applied in high density magnetic storage [15], magnetoelectric transducers [16] and spin filters [17]. Although the face-center-cubic (fcc) inverse spinel structure of CFO shares almost no similarity with the rhombohedral corundum structure of $Al_2O_3$, CFO films of (111) normal direction can be epitaxially grown on $Al_2O_3$(0001) substrates. The large difference in crystal structures and lattice parameter between these two materials and the large magnetoelastic effect of CFO implies the formation of interfacial layer of distinct magnetism, which is promising for large intrinsic EB.

In this work, we studied crystal structures, magnetism and electronic structure of CFO(111)/Al$_2$O$_3$(0001) thin films. We show that the apparent intrinsic EB, as large as 3 kOe, which resembles our more conventional understanding of EB, is the result of an antiferromagnetic layer adjacent to the ferromagnetic CFO layer. The formation of AFM CoO-dominant interfacial layers between CFO thin films and substrates are identified by structural, magnetic, and electronic structure measurements, which is the origin of the intrinsic EB and the anomalies in structural and magnetic data. Near-room temperature EB can be observed in CFO(111)/Al$_2$O$_3$(0001) thin films.

## Methods

**Sample preparation**. (111)-oriented CFO thin films of thicknesses from 1.7 to 55.4 nm were grown on $\alpha$-Al$_2$O$_3$(0001) substrates by pulsed laser deposition (PLD). The KrF excimer laser of wavelength 248 nm was employed to ablate the CFO target with a pulse energy of 140 mJ and a repetition rate of 2 Hz. The oxygen partial pressure was 10 mTorr during the growth. For all samples, the substrate temperatures were kept at 600 °C by a laser heater system during the growth. The whole growth process was in-situ monitored by a reflection high energy electron diffraction (RHEED) system. After the film growth, all the samples were annealed ex-situ in one-atmosphere oxygen gas at 600 °C for 5 hours.

**Structure characterization**. The out-of-plane $\theta$-$2\theta$ x-ray diffraction (XRD) and x-ray reflectivity (XRR) were conducted using a Rigaku D/Max-B x-ray diffractometer (cobalt K-α source, $\lambda$ = 1.793 Å) and a Rigaku SmartLab x-ray diffractometer (copper K-α source, $\lambda$ = 1.5406 Å), respectively; the film thickness was extracted from the XRR data. The in-plane crystal structure was studied by analyzing time-resolved RHEED patterns recorded every 30 seconds.

**Magnetic characterization with SQUID**. The magnetic hysteresis loops were measured in a superconducting quantum interfere device (SQUID) system with the cooling field +/- 70 kOe.

**X-ray photoelectron spectroscopy.** X-ray photoemission spectra were acquired using VG100AX hemispherical analyzer and using a SPECS X-ray Mg K$_\alpha$ anode ($hv$ = 1253.6 eV) source. All the XPS measurements were carried out at room temperature in an ultra-high vacuum chamber with a base pressure better than $3\times10^{-9}$ torr.

## Results

### Structural characterization and evidence of interfacial reconstruction

The bulk (inverse spinel) structure of CFO is illustrated in Fig. 1a using the unit cell. The fcc close-packed lattice of oxygen anions include two types of cation interstices: tetrahedral A sites with coordination 4 and octahedral B sites with coordination 6; with the cubic point symmetry, 1/8 of A sites are occupied by $Fe^{3+}$, while 1/2 of B sites with slightly distorted trigonal symmetry are occupied by $Fe^{3+}$ and $Co^{2+}$. The large structural difference between CFO and Al$_2$O$_3$ suggests a strong reconstruction at the interface of epitaxial CFO/Al$_2$O$_3$ films for the transition between the two structures.

Out-of-plane crystal spacings of the CFO films were studied using XRD $\theta$-$2\theta$ scan. A representative scan for a 22.5-nm-thick sample is shown in Fig. 1b, indicating no obvious impurity phases and the CFO (111) // $Al_2O_3$ (0001) epitaxial relation. The spacing between the (111) planes $d_{(111)}$, calculated from CFO (222) diffraction peak (see Supplementary Material Fig. S2), is displayed in Fig. 1c as a function of film thickness $t$. Except for the data point at $t$ = 3.33 nm, $d_{(111)}$ increases with $t$ with all values below the bulk value [18], demonstrating that the CFO films are under a tensile strain which is released gradually as the thickness increases. The outlier data point at $t$ = 3.33 nm suggests the existence of an interfacial layer with a different lattice spacing.

To tackle the structural details of the interfacial layer, in-plane crystal structures of the films were studied using in-situ RHEED. The typical RHEED images of both CFO thin films and $\alpha$-$Al_2O_3$ substrates are shown in Fig. 2a. The electron beam was along the CFO [1$\bar{1}$00] and [11$\bar{2}$0] directions for the left and right two images, respectively. The streaky patterns are consistent with the finite size of CFO grains and the 2-dimensional nature of diffraction, which indicate smooth film surfaces. Because of the fcc spinel structure of CFO, only all-odd and all-even Miller indices can survive. The in-plane epitaxial relation can be extracted from the RHEED pattern, as depicted in Fig. 2b, with a 30° rotation between the in-plane reciprocal unit cell of CFO relative to that of sapphire substrate, which is the same as that of the $Fe_3O_4$/$Al_2O_3$ films [19].

Time-resolved RHEED was carried out to elucidate the depth profile of the film by in-situ monitoring the top layer of the film during the growth [19]. RHEED images along the [1$\bar{1}$00] direction of CFO was taken every 30 seconds during the film growth at a repetition rate of 1 Hz. Given the growth speed ~1.34 Å /min, the deposition time can be converted to film thickness. The RHEED images were then summed up along the streaks direction to form RHEED spectra, i.e., RHEED intensity as a function of horizontal position. Combining all the spectra, one reaches the 2-dimensional representation of the time (or thickness) -dependence of RHEED pattern, as shown in Fig. 2c. When the thickness $t$ is less than 2 nm, the CFO (-6-60), (-2-20), (220), (660) diffraction lines (weak lines in the yellow dashed box) disappear, suggesting a structural reconstruction in the interfacial layer. To quantify this observation, intensity of the weak diffraction lines was calculated and compared with the relative in-plane lattice constant extracted from diffraction spacings as a function of film thickness; the results are shown in Fig. 2d. The in-plane lattice constant decreases with thickness, which is a direct evidence of the tensile strain, confirming the results of $d_{(111)}$ thickness dependence in Fig. 1c. The abrupt reduction of in-plane lattice constant at around 2 nm coincides with the abrupt change in diffraction intensity of the weak lines. The disappearance of the weak diffraction lines and the persistence of the stronger lines near the interface, suggest that the interfacial layer has a similar structure as CFO but with a lattice constant roughly half of that of CFO, because doubling the reciprocal (diffraction) spacing corresponds to halving the real-space spacing.

Therefore, the nature of the nominal CFO(111)/$Al_2O_3$(0001) epitaxial films can be represented by a three-layer structure, as illustrated in Fig. 2e. The thickness of interfacial layer is consistent with the thickness of "dead layer" in the CFO/$Al_2O_3$/Si (111) revealed using x-ray magnetic circular dichroism [20]. According to the lattice constants of the interfacial layer and the possible compositions of Fe and Co oxides, CoO and FeO, with fcc rock-salt structures with lattice constants $a_{CoO}$ = 0.42630 nm [21] and $a_{FeO}$ = 0.43108 nm respectively, are obvious candidates. To compare with the transition in Fig. 2d, we calculate the differences, $2a_{CoO} - a_{CFO} = 0.015$ nm and $2a_{FeO} - a_{CFO} = 0.024$ nm. Apparently, the lattice constant of CoO matches the observation in Fig. 2d better. Using $a_{CoO} = 0.42630$ nm, the outlier lattice spacing value at $t = 3.33$ nm (Fig. 1c) could also be obtained.

Magnetic characterization and colossal EB

Fig. 3a and b show the typical out-of-plane and in-plane hysteresis loops of an 8.6-nm-thick sample measured at 20 K. All the hysteresis loops were measured after the samples were cooled down under magnetic field +/- 70 kOe. The hysteresis loops of CFO(111) thin films demonstrate small out-of-plane magnetic anisotropy, which is different from in-plane magnetic anisotropy of the CFO(001) films [22]. Both out-of-plane and in-plane hysteresis loops indicate that the samples contain two magnetic components ("soft" component with a small coercive field and "hard" component with a large coercive field;) where the "hard" component corresponds to the CFO component. The magnetic-field bias on the hysteresis loops generated by the field-cool (FC) condition, i.e. the EB, is clearly observed in Fig. 3a. An analysis of the derivative $dM/dH$ shows that the contribution of the soft component to EB is negligible (see Supplementary Material Fig. S3). Based on the symmetry of the hysteresis loop and the thickness dependence of magnetic moment of the hard-component, the soft components can be subtracted (see Supplementary Material Fig. S3). The EB ($H_{EB}$) and coercivity ($H_C$) of the hard components can then be calculated for Fig. 3a and Fig. 3b, resulting in a colossal out-of-plane value $H_{EB} = 3.13$ kOe and an in-plane value $H_{EB} = 0.89$ kOe respectively. The out-of-plane EB observed here is comparable to the 3.65 kOe EB reported in the CFO-CoO core-shell nanoparticles [23], suggesting the interfacial similarity between two systems, consistent with the scenario of CoO interfacial layer between the CFO film and the $Al_2O_3$ substrate.

Fig. 2c and d show $H_{EB}$ and $H_C$ derived from the hysteresis loops measured using the same condition as that of Fig. 2a for films of different thicknesses. The thickness dependence of $H_{EB}$ can be fitted by a power law $H_{EB} \propto 1/t_{CFO}^{0.4}$, where $t_{CFO} = t-t_0$, $t_0 = 2$ nm is the thickness of the interfacial layer estimated from the RHEED analysis. The power-law thickness dependence of $H_{EB}$ indicates that the EB comes from an interfacial effect where the magnetization of the FM CFO is pinned by the AFM CoO layer by the exchange interaction. Instead of power 1 which corresponds to a sharp and ideal interface in all the interface-based EB models [24,25], the power 0.4 may originate from the finite transition thickness between the CoO layer and the CFO layer revealed in Fig. 2d.

The coercivity $H_C$ in general increases with $t_{CFO}$ and reaches saturation at about $t_{CFO} = 14.6$ nm, as shown in Fig. 4d. The increasing trend of $H_C$ when thickness increases ($t_{CFO} > 4.7$ nm), is consistent with previous results explained as more antiphase boundaries in thinner films [26,27]. Similar to the thickness dependent of $d_{(111)}$, the data point at $t = 3.33$ nm is an outlier that does not follow the overall trend, indicating a different nature of the FM layer at small thickness, which again may have to do with the finite transition thickness between CFO and CoO.

Temperature dependence of $H_{EB}$, $H_C$ and magnetization

The temperature dependence of $H_{EB}$, $H_C$ and saturation magnetization $M_S$ of the $t_{CFO} = 6.6$ nm CFO thin film for both out-of-plane and in-plane were plotted in Fig. 4. Both $H_{EB}$ and $H_C$ along out-of-plane and in-plane directions demonstrate similar unimodal trend in temperature, with the maximum values at approximately $T = 25$ K. While the $H_{EB}$ disappear at around $T = 250$ K, $H_C$ reaches their minima (about 1kOe) in the 250 – 300 K range. Although the Néel temperature of bulk CoO is about 290 K [28], the value for thin films may reduce due to the finite-size effects [29]. For example, the Néel temperature of a 2-nm-thick CoO layer may reduce to 250 K [29], which is consistent with the temperature range at which the out-of-plane and in-plane $H_{EB}$ disappear, as shown in Fig. 4a and b. In contrast, the AFM transition temperatures of other possible compounds (listed in Table S1) cannot fit the experimental observation.

Both the out-of-plane and in-plane values of $M_S$ stay around 1.5 μB/f.u. for $T >$ 25 K and abruptly increase at $T = 25$ K toward low temperature to about 4.5 μB/f.u. at $T = 5$ K. The sharp peaks of $H_{EB}$ and $H_C$ in Fig. 4a and b appear to echo with the sudden change of $M_S$ at around $T = 25$ K. If we take the interfacial EB model [24,25] and the Stoner-Wohlfarth model, both $H_{EB}$ and $H_C$ are inversely related with the average magnetization. That's exactly why they begin to sharply decease as $M_S$ suddenly increase at $T = 25$ K towards low temperatures, as indicated in Fig. 4a and b.

Electronic structural characterization and confirmation of interfacial layer

To further elucidate the nature of the interfacial layer, we studied the electronic structure of CFO(111)/Al$_2$O$_3$(0001) films using x-ray photoelectron spectroscopy (XPS). The XPS Co $2p_{3/2}$ core level spectra for CFO films of thicknesses 5.5 nm and 1.7 nm grown on Al$_2$O$_3$ are depicted in Fig. 5a and c, respectively. For the $t = 5.5$ nm film, the Co $2p_{3/2}$ XPS core level spectra contain three components: $P_1$ at 781.4 eV, $P_2$ at 783.7 eV, and S (satellite) at 788.1 eV. For the $t = 1.7$ nm film, these three Co $2p_{3/2}$ core level features, $P_1$, $P_2$, and S are at the smaller binding energies of 781.0 eV, 783.4 eV, and 786.5 eV respectively.

As illustrated in Fig. 5, for both the 1.7 nm and the 5.5 nm films, the binding energies of $P_1$, $P_2$, and S Co $2p_{3/2}$ core level components are somewhat greater than the CFO film Co $2p_{3/2}$ core level component binding energies of 780.4 eV, 782.8 eV and 786.2 eV [30] and 779.8 eV, 781.9 and 785.9 eV [31] reported previously. These somewhat larger binding energies are consistent with a dielectric CFO grown on a dielectric substrate, although less than the binding energies of 787.0 eV, 789.4 eV and 793.4 eV reported elsewhere [32]. For the 1.7 nm film, the binding energies of some of these three Co $2p_{3/2}$ core level features are in agreement with the results reported by Wan and Li [33]. Although the spectrum for Co $2p_{3/2}$ core level features in the work of Wan and Li [33] lacks a peak equivalent to $P_2$ in our work, the binding energy values of spectral components $P_1$ and S in our work for a 1.7 nm thick film are in agreement with their values of 780.9 eV and 785.5 eV respectively. The value of binding energy for the Co $2p_{3/2}$ core level feature (S) seems to be higher in our work, than in the work of Wan and Li [33] but this could be a result of a poor fitting of the Co $2p_{3/2}$ spectrum in the latter work.

These three Co $2p_{3/2}$ core level features, labeled $P_1$, $P_2$, and S in Fig. 5, are typically assigned to the cobalt placed in the cation octahedral and tetrahedral sites [30-32] as well as Co $2p_{3/2}$ core level satellite feature at even larger binding energy, respectively. The cation tetrahedral and octahedral sites are the A and B sites of Fig. 1a. The core level photoemission satellite feature is consistent with a 2-hole bound state common to oxides with a band gap. Nonetheless, the $P_1$ and $P_2$ core level features in the Co $2p_{3/2}$ XPS spectra for the 5.5 nm film and that of the 1.7 nm film are tantamount to a surface to bulk core level shift. This is evident in the changing ratio of $P_1$ to $P_2$ with emission angle, obtained from angle-resolved XPS on the 5.5 nm film as plotted in Fig. 5b, since XPS becomes more surface sensitive as the photoelectron emission angle, with respect to the surface normal, increases [34-36].

A surface-to-bulk core level shift in the core level binding energy is ascribed here to the different chemical environments of the Co cations at the surface compared to the Co atom in the bulk part of the 5.5 nm film [35-38]. Higher ratio of $P_1/P_2$ Co $2p_{3/2}$ XPS spectra components, at higher emission angles, show that more $P_1$ species is present at the surface than $P_2$, thereby making $P_1$ the surface core level and $P_2$ the bulk core level components of Co $2p_{3/2}$ core level. Such a change in $P_1/P_2$ ratio is consistent with a surface with a cobalt species that differs from the bulk and calls into question the assignments of octahedral $O_h$ and tetrahedral $T_d$ site occupancy based on the core level photoemission intensities, as done elsewhere [30-32].

In addition to the binding energy shifts in the Co $2p_{3/2}$ core level photoemission components, observed between 5.5 nm and 1.7 nm thick CFO films, the 1.7 nm film has an additional peak ($P_0$) with 779.4 eV binding energy, changing the shape of Co $2p_{3/2}$ core level spectra at the lower binding energies. This additional Co $2p_{3/2}$ core level binding energy component, $P_0$, has a larger binding energy than 778.3 eV, the Co $2p_{3/2}$ core level binding energy of cobalt metal [39] and is indicative of a reduced oxide in the Co $2p_{3/2}$ core level spectrum in the 1.7 nm film. Since the 1.7 nm film is close to the ~ 2 nm interfacial AFM layer discovered from the structural characterization, the electronic structure of this film is expected to reflect the properties of the CFO/$Al_2O_3$ interface. This additional $P_0$ Co $2p_{3/2}$ XPS spectra component, in the thinner CFO films, supports the scenario where there is a CoO interfacial layer. In addition, the smaller binding energies overall and the larger number of Co $2p_{3/2}$ XPS spectral components is indicative of suboxide $Co_xO$ (x>1) formation. In contrast, there are no additional peaks for the Fe $2p_{3/2}$ for CFO films of thicknesses 5.5 nm and 1.7 nm (see Supplementary Material Fig. S5).

**Discussion**

Magnitude of the saturation magnetization $M_s$ and possible magnetic transition

In the ideal ferrimagnetic bulk CFO, tetrahedral A sites are occupied by $Fe^{3+}$, while the octahedral B sites are shared by $Fe^{3+}$ and $Co^{2+}$. The $Fe^{3+}$ ions on A and B sites form colinear AFM order due to the 135° A-O-B superexchange interaction. The high-spin (3/2) $Co^{2+}$ ions on B-site form FM order, generating the net magnetization. The orbital angular momentum of the B-site $Co^{2+}$ is not quenched completely and the resultant spin-orbit coupling causes a large single-ion cubic magnetocystalline anisotropy of easy axes <100> [40]. The spontaneous spin and orbital moments of $Co^{2+}$ at 0 K was calculated as about 2.9-3 $\mu_B$ and 0.4-0.6 $\mu_B$, respectively [41]. Experimentally, the saturation magnetization were measured as 3.3-3.9 $\mu_B$/f.u. [42].

However, in CFO thin-film systems, reduced magnetizations varying from 0.47 to 3.48 $\mu_B$/f.u. were observed and several mechanisms were discussed [22]. In our systems, a room-temperature saturation magnetization value around 1.3 $\mu_B$/f.u. was measured (Fig. 4c and d). We can resort to the switching of high-spin ($3d^7$, $t_{2g}^5e_g^2$, $S=3/2$) $Co^{2+}$ to low-spin ($3d^7$, $t_{2g}^6e_g^1$, $S=1/2$) $Co^{2+}$ as a result of the tensile strain that distorts and strengthens the cubic crystal fields. Another possibility is that because the [111] tensile strain decreases the size of cobalt octahedra, $CoO_6$, large-size $Co^{2+}$ could be converted to smaller-size $Co^{3+}$ which normally takes a low-spin state ($3d^6$, $t_{2g}^6 e_g^0$, $S$=0) in octahedral sites [43] and meanwhile equal amounts of B-site $Fe^{3+}$ ($3d^5$, $t_{2g}^3 e_g^2$, $S$=5/2) turn to $Fe^{2+}$ ($3d^6$, $t_{2g}^4 e_g^2$, $S$=2). Both two low-spin cobalt mechanisms lead to a total spontaneous moment of 1 $\mu_B$/f.u.

Presumably, some structural phase transition related to the bonding between iron cations and oxygen anions or (and) double exchange between $Fe^{3+}$ and $Fe^{2+}$ [44] may occur at around $T$ =25 K. Since the low-temperature value of $M_S$ can be as large as 4.5 $\mu_B$/f.u, it seems impossible for cobalt ions to be the only main magnetic sublattice anymore. The assumed structural phase transition might turn B-site $Fe^{3+}$ or $Fe^{2+}$ into low-spin states or switch part of B-site cobalt ions with A-site iron ions, both of which could enable the total spontaneous moments arrive at 4.5 $\mu_B$/f.u or higher.

### Soft component in the hysteresis loops

The anomaly of the thickness-dependent $H_C$ at $t$ = 3.33 nm could be ascribed to the $Fe_3O_4$-like component $Co_{1-\delta}Fe_{2+\delta}O_4 (\delta \lesssim 1)$ which corresponds to the magnetic soft component in hysteresis loops. $Co_{1-\delta}Fe_{2+\delta}O_4 (\delta \lesssim 1)$ could exist as the transition layer between the CoO layer and the well-defined CFO layer; its thickness can be estimated by that of the transition layer, i.e., about 1 nm as demonstrated in Fig. 2d. From the hysteresis loop of the $t$ = 1.7 nm sample (see Fig. S3c), the estimated magnetization of $Co_{1-\delta}Fe_{2+\delta}O_4$ is larger than 8 $\mu_B$/f.u. This extraordinary large moment reveals that, as a result of strong interfacial reconstruction, its cation distribution and structure may be significantly different from that of ideal inverse spinel whose spontaneous moment is smaller than 5 $\mu_B$/f.u.

### Conclusions

We have investigated the structure, magnetic properties and electronic states of CFO(111) /Al$_2$O$_3$(0001) films. A large intrinsic EB was measured when the magnetic field is applied in both the out-of-plan and the in-plane direction. The appearance of EB and the experimental data around the interface between the Al$_2$O$_3$ substrate and the CFO thin film indicate that a CoO-$Co_{1-\delta}Fe_{2+\delta}O_4 (\delta \lesssim 1)$ interfacial layer has formed due to the interfacial induced compositional reconstruction and stoichiometry changes. As such, here the intrinsic EB looks very much like conventional EB as there is an antiferromagnetic layer in proximity to ferromagnetic CFO layer. A strain-induced low-spin cobalt ion is proposed to explain the low magnetization in this system and a potential structural phase transition emerged at around $T$ = 25 K. A surface-to-bulk shift in the core level binding energy in the XPS data further reveals the pure CFO thin film contains two components in electronic states. This work suggests a potential for engineering intrinsic EB, highlighting the role of atomic and electronic interfacial reconstruction.

### Acknowledgements

This work was primarily supported by the National Science Foundation (NSF), Division of Materials Research (DMR) under Grant No. DMR-1454618. The research was performed in part in the Nebraska Nanoscale Facility: National Nanotechnology Coordinated Infrastructure and the Nebraska Center for Materials and Nanoscience, which are supported by the NSF under Grant No. ECCS- 2025298, and the Nebraska Research Initiative.

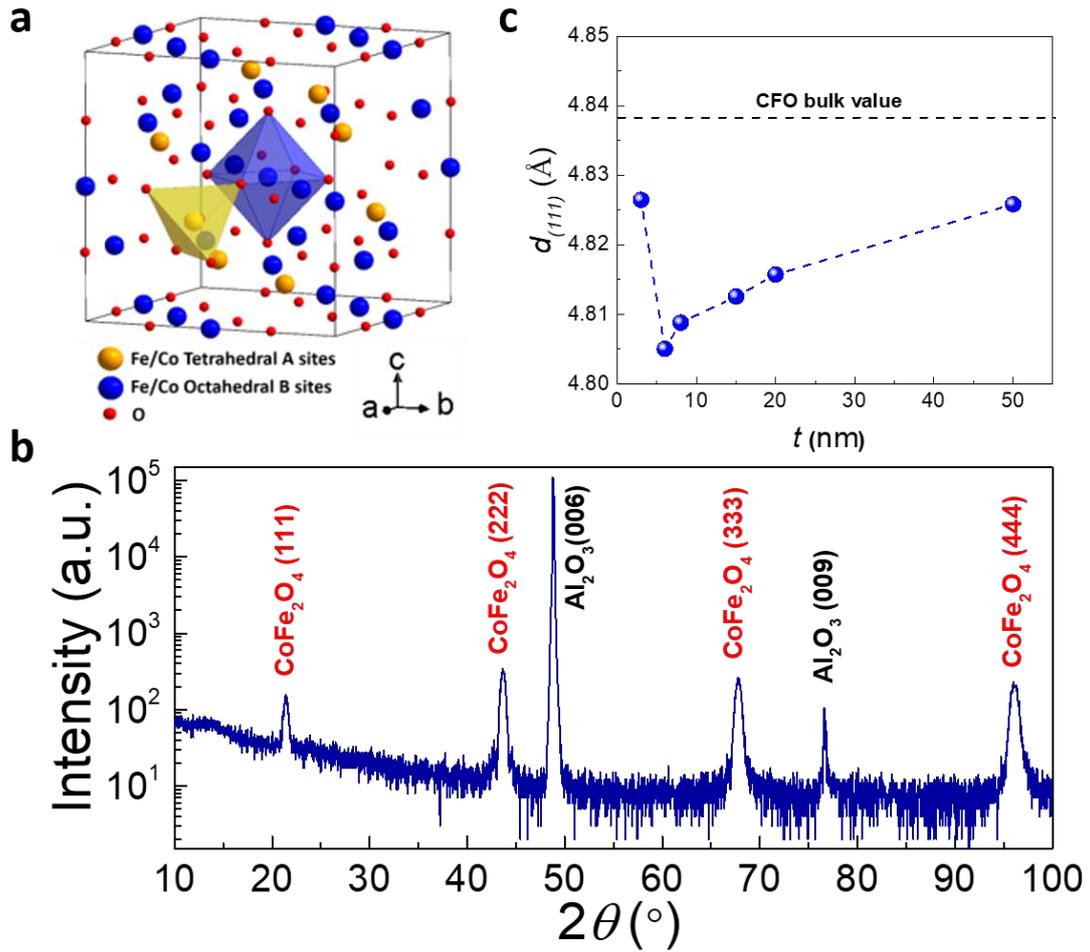

**FIG. 1. Structure characterization.** (**a**) Crystal structure of inverse spinel CFO, polyhedral model showing two interstitial sites: tetrahedral A sites marked by yellow and octahedral B sites marked by blue, respectively. (**b**) The $\theta$-$2\theta$ XRD results of a CFO film with a thickness of ~22.5 nm. (**c**) Thickness-dependent interplane spacing of CFO(111) lattice planes, indicating a tensile strain released with the increasing of thickness.

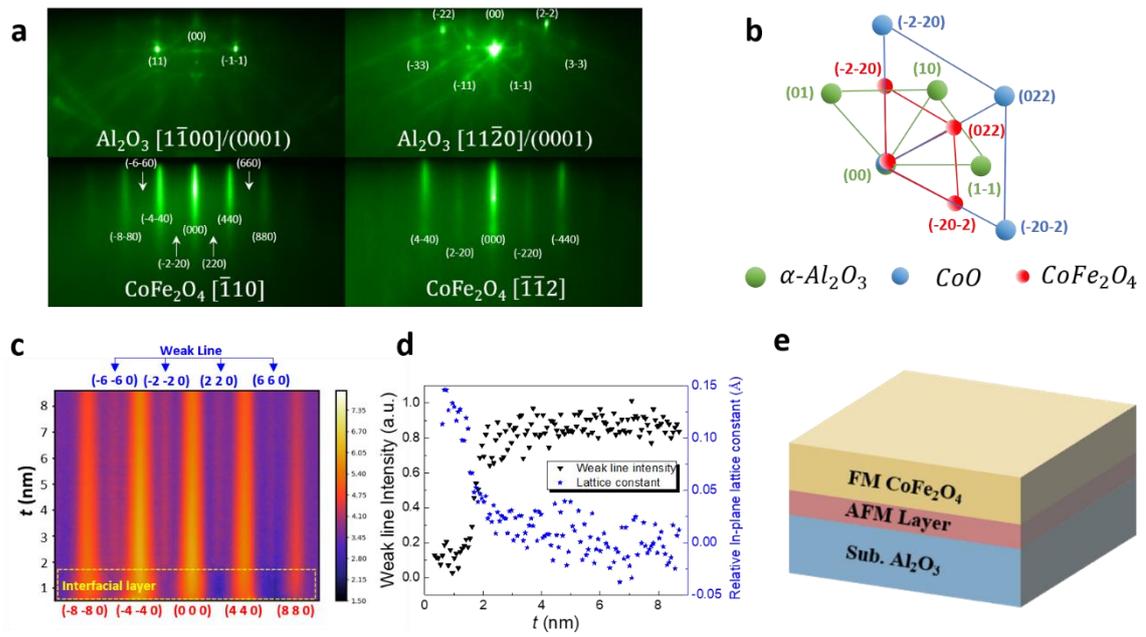

**FIG. 2. Epitaxial relationship of CFO/Al$_2$O$_3$ films and Time-resolved RHEED. (a)** RHEED pattern of CFO films and Al$_2$O$_3$ substrates along two perpendicular in-plane orientations: left [1$\bar{1}$00] and right [11$\bar{2}$0]. **(b)** in-plane reciprocal primitive unit cells of sapphire substrate (green), CoO (blue) and CoFe$_2$O$_4$ (red); the diffraction streaks are marked using their corresponding reciprocal indices projected into the (111) plane. **(c)** The evolution of [1$\bar{1}$00]-direction RHEED pattern during the growth. **(d)** Thickness dependence of weak-line intensity and relative in-plane lattice constant extracted from RHEED pattern. **(e)** Schematic diagram of a structure model of CFO/Al$_2$O$_3$ heterostructures.

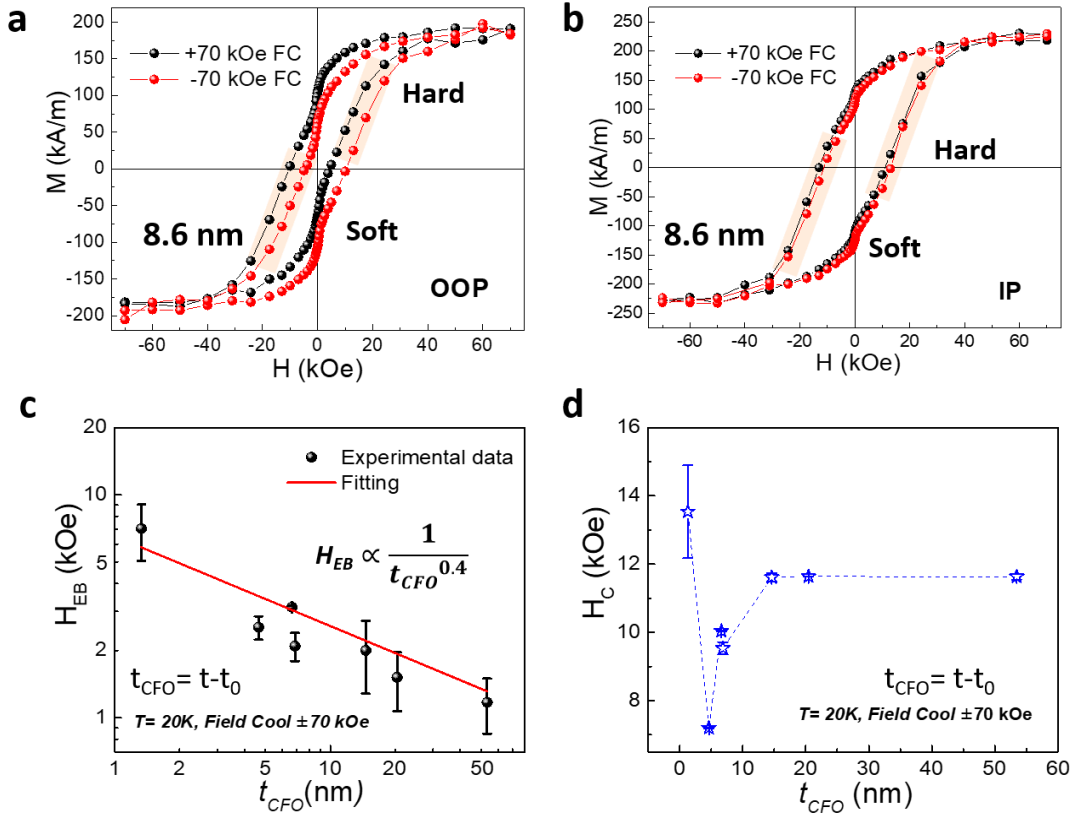

**FIG. 3. In-plane and out-of-plane EB**. Representative hysteresis loops of EB with cooling magnetic field of +/- 70 kOe along **(a)** out-of-plane direction and **(b)** in-plane direction. Measuring temperature is 20 K, and the thickness of this sample is 8.6 nm. After subtracting the soft component, the out-of-plane and in-plane EB read as 3.13 kOe and 0.89 kOe, respectively. Thickness-dependent EB, $H_{EB}$ **(c)** and coercivity field, $H_C$ **(d)** at 20K. In (c), black balls denote experimental data, while red line is the fitting curve. $t_{CFO} = t-t_0$, $t_0 = 2$ nm.

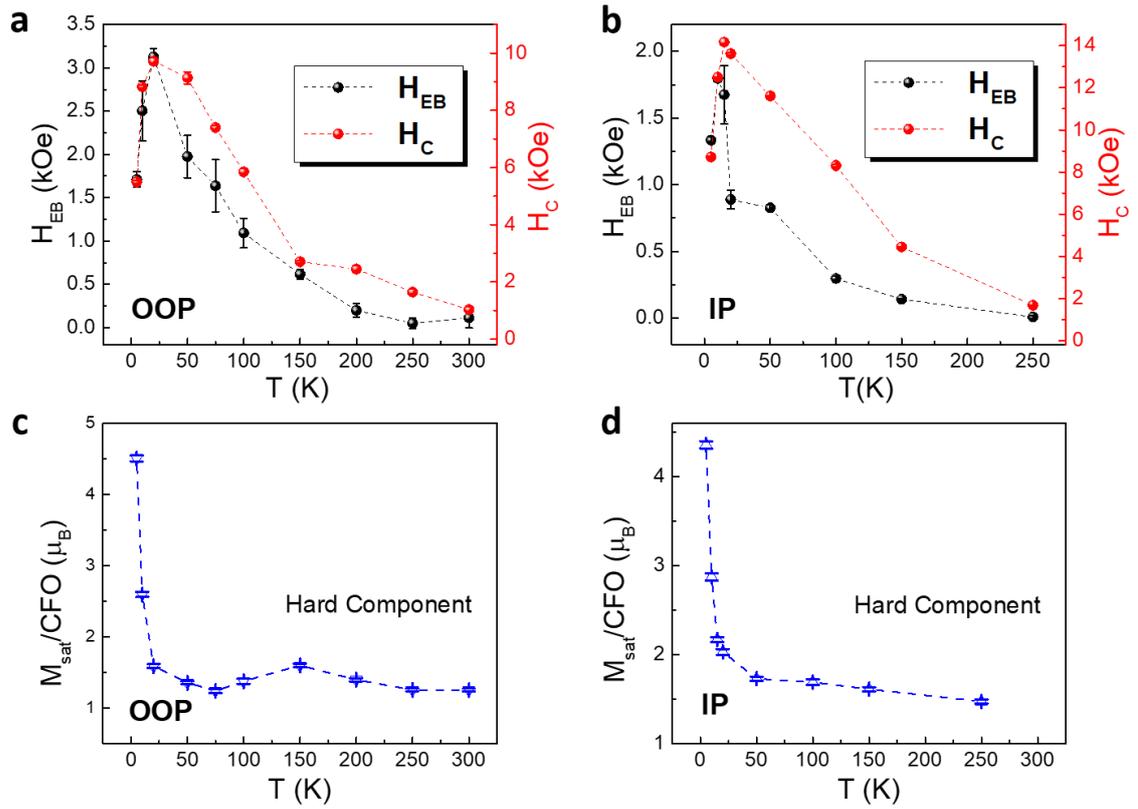

**FIG. 4. Temperature-dependent EB.** $H_{EB}$ and $H_C$ for the CFO thin film of 6.6 nm for out-of-plane **(a)** and in-plane **(b)**; temperature behaviors of saturation moments $M_S$ for the 6.6-nm-thick CFO thin film for out-of-plane **(c)** and in-plane **(d)**.

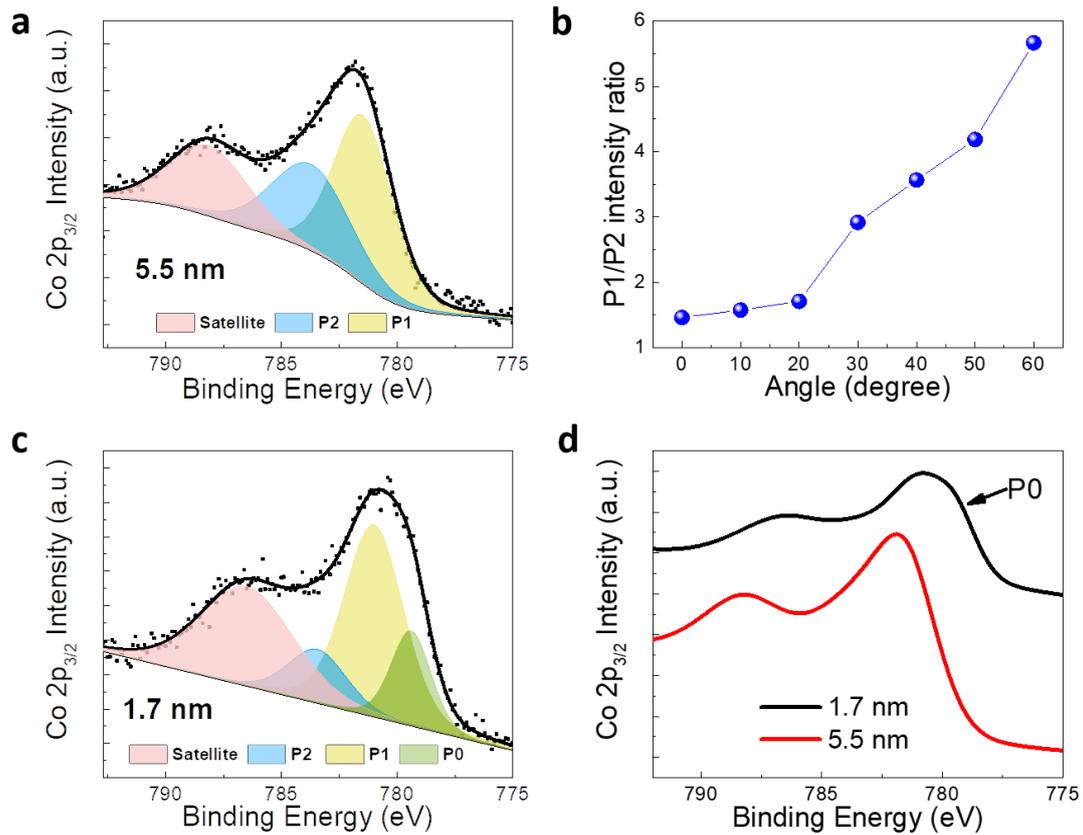

**FIG. 5. XPS of CFO/Al$_2$O$_3$ films.** The XPS of the Co 2p$_{3/2}$ core level features in CFO/Al$_2$O$_3$ with the CFO film of thicknesses **(a)** 5.5 nm and **(c)** 1.7 nm. The Co 2p$_{3/2}$ core level photoemission spectrum, for a 5.5-nm-thick film, contains three peaks: P$_1$, P$_2$, and Satellite (S) **(a)**. An additional peak (P$_0$) on the lower binding energy side to P$_1$ is observed in the 1.7-nm-thick film **(c)**. **(b)** The P$_1$/P$_2$ XPS component intensity ratios for the Co 2p$_{3/2}$ core level, from 5.5-nm-thick film, are plotted as a function of the photoemission take-off angle with respect to the surface normal. (d) The fitted spectra of Co 2p$_{3/2}$ core levels for the films of thicknesses 5.5 nm and 1.7 nm, displaying clear evidence of the dielectric nature of the thicker CFO films and of an additional component peak (P$_0$), in the thinner films.